\documentclass[sigconf]{acmart}

\usepackage{multirow}
\usepackage{siunitx}
\usepackage{xspace}
\usepackage{enumitem}%
\setlist[itemize]{noitemsep, topsep=0pt}
 \usepackage{balance}
 \usepackage[T1]{fontenc}
\usepackage[utf8]{inputenc}

\AtBeginDocument{%
  \providecommand\BibTeX{{%
    \normalfont B\kern-0.5em{\scshape i\kern-0.25em b}\kern-0.8em\TeX}}}

\copyrightyear{2021}
\acmYear{2021}
\setcopyright{acmlicensed}\acmConference[ITiCSE 2021]{26th ACM Conference on Innovation and Technology in Computer Science Education V. 1}{June 26-July 1, 2021}{Virtual Event, Germany}
\acmBooktitle{26th ACM Conference on Innovation and Technology in Computer Science Education V. 1 (ITiCSE 2021), June 26-July 1, 2021, Virtual Event, Germany}
\acmPrice{15.00}
\acmDOI{10.1145/3430665.3456370}
\acmISBN{978-1-4503-8214-4/21/06}

\begin{document}
 \fancyhead{}

\title{Novices' Learning Barriers When Using Code Examples in Open-Ended Programming}

\author{Wengran Wang, Archit Kwatra, James Skripchuk, Neeloy Gomes, Alexandra Milliken,\\ Chris Martens, Tiffany Barnes, Thomas Price}

\affiliation{%
  \institution{North Carolina State University}
  \city{Raleigh}
  \state{NC}
  \country{USA}
}

\newcommand{\q}[2]{"\textit{#1}"{#2}}

\newcommand{\bq}[2]{\vspace{5px}\noindent\textit{#1}~{\small [P#2]} \vspace{5px}}
\newcommand{\snap}{Snap\textsl{!}}
\renewcommand{\quote}{\list{}{\rightmargin=\leftmargin\topsep=2.5pt}\item\relax}

\renewcommand{\shortauthors}{}

\begin{abstract}
Open-ended programming increases students' motivation by allowing them to solve authentic problems and connect programming to their own interests. 
However, such open-ended projects are also challenging, as they often encourage students to explore new programming features and attempt tasks that they have not learned before. Code examples are effective learning materials for students and are well-suited to supporting open-ended programming. However, there is little work to understand how novices learn with examples during open-ended programming, and few real-world deployments of such tools. In this paper, we explore novices' learning barriers when interacting with code examples during open-ended programming. We deployed Example Helper, a tool that offers galleries of code examples to search and use, with 44 novice students in an introductory programming classroom, working on an open-ended project in \snap. We found three high-level barriers that novices encountered when using examples: decision, search, and integration barriers. We discuss how these barriers arise and  design opportunities to address them.

\end{abstract}

\maketitle

\section{Introduction} 
Creative, open-ended programming projects, such as making student-designed apps, games and simulations, are widely used in many introductory programming courses (e.g. \cite{garcia2015beauty}). They encourage novices to pursue projects that feel authentic to them, and to express their ideas creatively, motivating them to keep pursuing CS \cite{guzdial2005design}. In addition, through open-ended programming, novices also learn to use computational thinking strategies (e.g., abstraction, decomposition), and may further apply them in other areas, such as math and engineering \cite{urban2000there, guzdial2003computer}. 


Open-ended projects can prove very difficult for novices \cite{grover2018what}, in part because they encourage novices to design unique programs that address their interests and goals, which may require them to use new programming features, or accomplish new tasks, beyond what they have already learned. Novices can also struggle to combine the \textit{individual} programming concepts they have learned (e.g. loops, variables, etc.) into a complete program \cite{grover2018what}, and they may lack experience making use of code blocks or libraries offered by the language (i.e. APIs \cite{gao2020exploring}).

Code examples are a common way for programmers to learn new APIs and coding patterns \cite{brandt2009two}, and are also considered one of the most useful learning material for novices \cite{lahtinen2005study}. For example, research in laboratory settings suggests that novices learned to use code blocks more effectively after seeing them from code examples \cite{ichinco2017suggesting}. However, novices can also face challenges learning from examples, and integrating examples to their own code \cite{ichinco2015exploring}, and these challenges may be exacerbated by the challenges of open-ended programming \cite{ichinco2019open}. In addition, there have been few real-world deployments of code examples for supporting open-ended programming. To design example systems that better support novices' open-ended programming, a key step is to uncover their own barriers and frustrations \cite{guzdial2015learner}. This suggests the need to explore how novices use code examples in practice, especially in a \textit{classroom} setting, with authentic population and learning activities.

In this work, we ask the research question: \textbf{What are the learning barriers that novices face when using examples during open-ended programming?}. To answer this, We designed a system called Example Helper to support open-ended programming with a gallery of code examples. Our analysis of log and interview data found that novices encounter three types high-level barriers: decision, search and integration barriers. Based on these findings, we discuss implications and design opportunities for better supporting novices' open-ended programming with examples. The primary contributions of this work are: 1) The Example Helper system that offers a variety of learning support to novices during open-ended programming. 2) An analysis of learning barriers novices encounter when using code examples in open-ended programming, in an authentic, classroom context. 3) Identification of design opportunities to provide better example-based support to novices.

\section{Related Work}
\noindent\textbf{Open-ended programming.}
~\\
Open-ended programming allows learners to integrate personal interests into creating an artifact that is meaningful to them. Many efforts to promote open-ended programming draw on the theory of Constructionism \cite{papert1980mindstorms}, which suggests that learners effectively build their own knowledge structure when engaging in creating a programming artifact they feel connected with \cite{papert1980mindstorms}. 
\par
However, open-ended projects can be challenging for novices. 1) They may struggle to design ``logically-coherent'' programming components, and may start by putting together all possible code elements that seemed relevant \cite{Meerbaum-Salant2011}. 2) Their programs may suffer from code smells such as duplicated code \cite{robles2017software}. 3) Their final artifacts were shown to be lack of usage of fundamental programming concepts (e.g., variables, operations), from a systematic evaluation of 80 novices' open-ended projects collected from 20 urban middle school classrooms \cite{grover2018what}. These show struggles to apply existing concepts into code, or to explore new programming concepts or APIs. \citeauthor{kirschner2006minimal} summarized through a literature review that open-ended discovery may lead to experiential learning, where learners rely heavily on trial-and-error instead of learning new knowledge \cite{kirschner2006minimal}. These challenges encountered by novices during open-ended programming are examples of ``Play Paradox'' \cite{noss1996windows}, which explains that learning activities should strike a balance between creative exploration and some levels of external support \cite{noss1996windows}.

\noindent\textbf{Code examples \& opportunistic learning.}
~\\
\noindent\textit{Theory.}
Code examples are one of the primary resources professional programmers and end-users use to learn programming knowledge and API usage patterns \cite{robillard2009makes,brandt2009two}. Such an example usage scenario arises when a programmer feels in need of resources in the middle programming. They search for a code example (e.g., through documentation or forums) \cite{brandt2009two}, and then integrate the example to their project through testing and modification \cite{brandt2009two}. Prior work has shown that, different from learning traditional Worked Examples \cite{clark2016learning}, where programmers engage in \textit{deliberate learning} of a step-by-step demonstration \textit{before} working on the actual task \cite{morrison2015subgoals, trafton1993, pirolli1994learning}, learning an example \textit{in the middle of} programming is a type opportunistic learning \cite{brandt2009two, gao2020exploring}, where programmers search, select, and copy code examples to ``get something to work with'', and then briefly test or modify to integrate examples into their own code \cite{rosson1993active}. When investigating experienced programmers' opportunistic learning, \citeauthor{rosson1993active} found that these programmers made effective use of examples to complete functionalities that they were unfamiliar with, but many don't reflect on \textit{how} the example works \cite{rosson1993active}. They may also struggle to apply or extend examples afterwards \cite{thayer2020thesis}. While this explains the experts' opportunistic learning of code examples, and described \textit{how} experts can encounter difficulties in using and applying code examples, it is unclear how this theory will extend to novices.
\par
\textit{Systems.} Researchers have developed systems to support novices' use of code examples during programming. Many were built for closed-ended tasks \cite{wang2020iticse, zhi2019sigcse}. For example, by offering step-by-step examples with options to immediately run the example code \cite{wang2020iticse}. Some offers an online database of annotated examples \cite{brusilovsky2001webex}. Some prior work has shown that novices benefited from requesting such code examples. For example, \citeauthor{ichinco2017suggesting} designed examples with options to contrast an alternative choice, and found novices used code blocks more effectively after seeing them from examples \cite{ichinco2017suggesting}. While this suggests that novices may potentially benefit from using code examples during programming, there are few real-world deployment of example-based tools to support novices' \textit{open-ended programming} in particular.  

Some prior work analyzed novices' learning barriers when using these example-based support systems \cite{ichinco2015exploring}. \citeauthor{ichinco2015exploring} analyzed novices' example use in Looking Glass, and found that they struggled to understand examples, and may fail to connect an example to their own task \cite{ichinco2015exploring}. However, this particular study explored novices' learning barriers in \textit{closed-ended} code completion tasks during a lab study. In contrast, we investigate novices' learning barriers of using code examples in \textit{open-ended} tasks, where students can encounter a different set of challenges.

\section{Example Helper System}


\begin{figure}
   \centering
   \includegraphics[width=.49\textwidth]{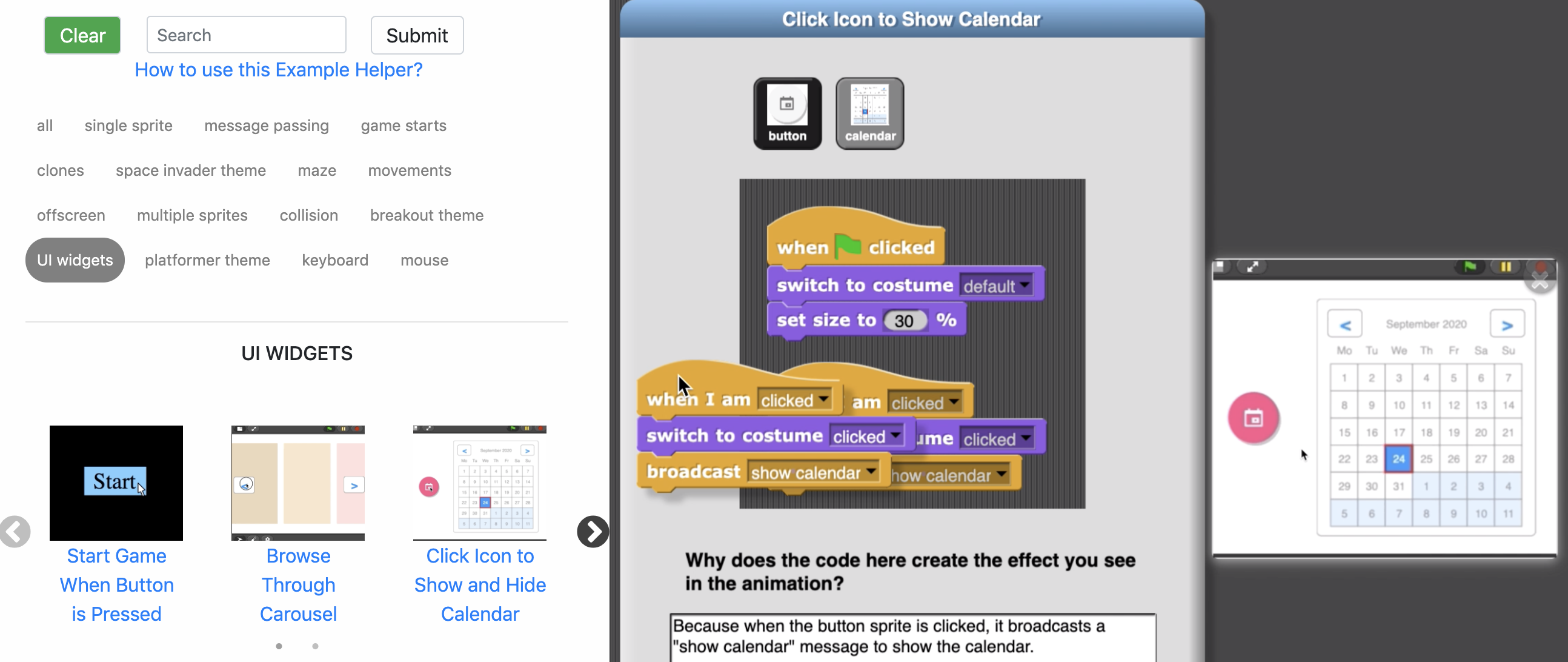} 
   \caption{The Example Helper Interface.}
   \label{fig:ui}
  \vspace{-4ex}
\end{figure}

The design goal of Example Helper is to allow students to view and incorporate existing programming patterns into their own code through effective use of code examples. To lower the barrier for making these programming projects \cite{ morelli2011can}, the system is incorporated into \snap~\cite{harvery2012byob}, a novice programming environment. Similar to other novice programming environments (e.g., Scratch\cite{resnick2009scratch}), \snap~already offers open-source galleries of complete programming artifacts from other programmers, but these are complete projects which demonstrate many related programming features. By contrast, Example Helper offers small snippets of code examples \cite{robillard2009makes} that demonstrate specific functionalities, collected in a curated, browsable gallery. We developed this curated set of examples through an analysis of students' programs from prior semester, extracting key program features that were shared across students, and built these as examples. Many of these key features include usage of multiple sprite interactions\footnote{A sprite in \snap~ is an object (i.e., in object-oriented programming) that has its own code (scripts), costumes (e.g., a button), and variables.} (e.g., in a collision event), we therefore also included examples that include usage of multiple sprites. Two experts then reconstructed examples from this repository to include cleaner and higher-quality code. When a student needs an example during programming, they can click on a ``show example'' button within the scripting area of \snap~to open a gallery of code examples. The student then follow two steps to select and use an example within their own source code: 

\noindent\textbf{Step 1: Search for an example.}
The student can find an example by: browsing through the gallery; or filtering and search for examples by clicking on a tag, or querying in a search box. The search box finds a set of examples the student need by looking for words that overlapped in the examples' names. To visually understand the functionality of the example, the student may also hover on the example to look at the gif animation of the code's output.

\noindent\textbf{Step 2: Use an example.} After finding a needed example, the student can click on the gif animation, and learn the example using the following steps:

\textit{Read the code in relation to the output.} The student may click on different sprites to look at the example code for each sprite (shown in Figure~\ref{fig:ui}). They may also look at the animation of the output next to the example code, since reading code in relation to output has been shown to trigger students to reflect on how the example code works \cite{wang2020iticse}. The student can also click on the ``Open the Project'' button to view the example in a separate window and experiment with it.

\textit{Write a self-explanation.} The student can reflect on the example by writing down a self-explanations: ``What in the code here creates the effect that you see in the animation?''. We designed self-explanation prompt because self-explanation is a critical step towards learning from an example \cite{price2020chi, atkinson2003}, since it promotes students to stop and think deeper about the code example \cite{aleven2016, gerjets2004}.

\textit{Copy the example code}. To allow students to test and modify the example easily, after writing their self-explanation of the example, the student may then drag and copy the example to their own code. To discourage students from immediately copying the code without thinking about it, we restricted the length of the self-explanation answer to be at least 30 characters.

\section{Participants \& Procedure}

%

We conducted our study in an undergraduate CS0 classroom for non-CS-majors with no prior programming experience, with 44 consented novice students, in a research university in Southeast US. The course was held online due to the COVID-19 pandemic. To create an authentic learning experience for the students, we did not collect their demographic information.

Students created open-ended projects over 3 weeks, starting from the 7th week of the course. Prior to that, they have learned the usage of fundamental programming concepts in \snap, including loops, conditionals, procedures, and lists. During the first week of project-making, students were introduced to the engineering design process \cite{haik2018engineering}, and were asked to make project pitches that may solve a real-world problem, including innovative ideas and user experience considerations. 

\noindent\textbf{Pair planning and programming.} Students discussed their project pitches online, and then may optionally form a 2-person group if they had a similar project of interests. 18 students chose to work individually, while the rest (26) worked in pairs, creating 31 student groups\footnote{Since some students worked alone and some in pairs, we use the term ``group'' to refer to the student or students who worked on a single project.}. Students then planned their project design in a digital planner \cite{milliken2021planit}. Before students started programming, one researcher came to the Zoom classroom and introduced the Example Helper. We also instrumented \snap~ to allow student pairs to easily transfer files through saving and loading, and encouraged them to use Zoom's screen share to collaboratively program. We encouraged students to collaboratively program because prior work has shown that in making open-ended programming projects, students achieved significantly higher performance in pair-projects than individual projects \cite{grover2018what}. 

\noindent\textbf{Interviews.} During the second week of project-making, we recruited 5 students to attend individual interview sessions with two researchers, where we recorded audio and  students' screens. During these interviews, we asked students: ``Is there anything you want to program, where you think an example might help you?'', and encouraged them to use an example and complete the feature during the interview. During this programming process, we asked students to think aloud \cite{greene2018capturing}. When they asked questions, we first encouraged them to think independently, and then offered them some possible next steps if needed. After completing the feature they wanted using the example, we asked about their experience using the examples, both during the interview and in their project-making experience, such as: ``Did you experience any difficulties using the examples?''. 

\section{Analysis}
\noindent\textbf{Qualitative Interview Data Analysis.}
To investigate our research question about students' barriers using code examples, we began by analyzing the interview data using thematic analysis \cite{braun2012thematic}. Two researchers each read thoroughly all interview data, and then individually conducted line-by-line inductive open coding on the five pieces of interview, to take note of any quotes or students' programming activities, that reflects their example-usage experience and their perceptions of it. While doing the inductive coding, the two researchers used each sentence as a segment, allowing 0 or more codes per segment. To obtain accurate understanding of students' experience during open-coding, they also used the screen-recording when students did the programming portion of the interview. The two researchers then discussed and resolved discrepancies. This created a merged set of 103 initial codes. The two researchers then investigated the 103 codes to identify ones that described students' learning barriers, and combined codes that described similar incidents of a type of barrier. This created 7 initial themes of learning barriers. They then discussed and sorted themes that may belong to a higher-level category, which created 3 high-level themes that described students' learning barriers, including 4 sub-themes.

\noindent\textbf{Log Data Preparation.}
\label{sec:log_analysis}
Based on the inductive and in-depth analysis on interview data, we discovered potential learning barriers among a small set of students. We then used log data to validate how these learning barriers are reflected across all participants, throughout their entire project-making classroom experience. Our log data included a total of more than 200 hours of programming activities (e.g., grabbing or destroying blocks), and students' code snapshots at every timestamp when they made a change to their code.
To elicit clean data that may be analyzed further to uncover novices' example-usage barriers, we performed a pre-filtering and prepared the following three types of the log data:

\textit{Search queries.} We collected all search queries that students have typed in the search box to look for an example.

\textit{Opened examples.} We manually investigated and then built a profile of each incident when a student opened an example, including: 1) {What examples were opened.} 2) {How the students found the example (e.g., whether they opened the best matches found by their search query). } 3) {What (if any) they did to integrate the example code to their own code (e.g., how they built, modified, or tested the example code).} 

\textit{Project submissions.} We analyzed students' final submissions to determine:  1) whether their project submissions included functionality demonstrated by the examples, and 2) whether the functionality came from their integration of a opened example, or from students' implementing the behavior independently.

Using the above filtered data, we further conducted deductive log data analysis based on the 7 themes collected from the interview, to find evidence of how these 7 learning barriers occurred in log data of all students, described in Section~\ref{sec:results}.

\section{Results \& Discussion}
\label{sec:results}

Our thematic analysis of the interview data revealed 7 barriers that students encountered when using code examples during open-ended programming, including 3 high-level categories: \textit{decision}, \textit{search}, and \textit{integration} barriers. For each barrier, we report our data by presenting the results from thematic analysis, and then the log analysis we conducted that may explain \textit{how} this barrier occurred in all 44 students. At the end of each barrier, we briefly discuss how this barrier relates to prior work, as well as its design implications.

\vspace{1mm}
\noindent \textbf{Decision Barrier: }\textit{Should I ask for an example?}

Our thematic analysis revealed that students encountered \textbf{decision barriers}, which occurred when students did not recognize their need or ability to ask for an example, even when they were stuck at implementing a programming behavior. For example, students may not consider asking for an example as an option: \textit{``[My partner] hadn't figured out how to implement a timer. I don't know why we didn't think about doing examples, but we didn't.''} (P3). 

Among 31 student groups, 27 (87.1\%) clicked on the ``show example'' button at least once to browse or search for an example, suggesting almost all were at least aware of the examples. However, we also found that 22\% of these students (6/27) opened the example interface only 1-2 times. This may suggest that students forgot about examples once they got started with their work, or the examples were not salient as they worked.

Another explanation could be that students judged the examples to be unhelpful after viewing the interface. While this may be the case for some students, we found that those who \textit{did} open the interface more than 2 times did so in an average of 11.38 times (up to 11 times for one group), suggesting that many students found it useful. We also found that 3 groups who did not use examples implemented functionality demonstrated by an example, totaling 7 times, suggesting examples would have been useful.

\textit{Discussion. } 
Prior work on novices' help-seeking behaviors has shown that knowing the need to seek help is an important but challenging self-regulatory skill that requires cognitive competencies \cite{karabenick2013}. Avoiding to seek help when stuck is a maladaptive learning strategy that can lead to reduced learning outcomes \cite{aleven2006toward}. This can be a particular challenge in programming, where students may have a strong desire to work independently, or get absorbed in their work and forget about asking for help \cite{price2017icer}. Our results suggest that this help-avoidance behavior also applies to novices' example use during open-ended programming. One possible way to address the problem of help avoidance in example systems like Example Helper is to offer help automatically (e.g. with a pop-up), which can reduce help avoidance \cite{marwan2020unproductive}, especially if the system can detect when students are stuck.

\vspace{1mm}

\noindent \textbf{Search Barrier: }\textit{How do I explain the example I want?} 

We found that only 63.0\% (17/27) of groups who clicked on the ``show example'' button ended up \textit{opening} a code example to view. Our thematic analysis suggests that this may have been the result of \textbf{search barriers}, where students sought an example but were unable to find or articulate what they were looking for. For example, \textit{``I think we had tried to look for a background that was like a sky or like a stage... and I don't believe we found one of those.''} (P2)

The log data reveals \textit{how} search barriers occurred in students' search queries. We found 63 distinct searches across the 15/27 groups (48\%) who used the search box to find examples(merging consecutive, identical queries). Two researchers conducted two rounds of coding on the queries to: 1) identify candidate themes that describe at least 10\% of the data, discuss to resolve conflicts; and 2) count the number of occurrences of each theme.  We found three primary themes: 1) interactions between multiple sprites, such as ``lose a point when touching'' or `shoot'' (14.3\%, 9/63); 2) sprite movement such as ``bounce'', or ``wrap around the screen'' (30.2\%, 19/63); 3) queries for how a sprite should look (rather than what it should do), such as ``dining room'',  ``airplane'', and  ``people'' (47.6\%, 30/63). While almost half of searches were in this category of how the sprite should \textit{look}, the examples were designed to show \textit{functionality}, so these searches returned no results -- such that only 39.7\% (25/63) of \textit{all} example searches yielded results. This shows a disconnect between how students \textit{articulated} the example they were looking for, based on aesthetic properties, and how examples are typically organized -- leading to search barriers.

Encountering a search barrier may also deter students from looking for examples in the future. Students who found and opened at least one example ($n=17$) used the ``show example" button over 5 time more (avg = 13.2; SD = 10.5) than those who did not (avg = 2.6; SD = 0.91).

In the interview, students discussed that they avoided requesting for help because of expectations that they won't find an example they needed:  \textit{``When I didn't find [a needed example], I kinda just steered away from [requesting examples].''} (P1), 

\noindent\textit{Discussion. }
Prior work on end-users' example search behaviors showed that they may not know how to articulate what it is they want to see in examples \cite{dorn2013lost}. Our analysis found similar results, that novices may also encounter difficulties expressing the \textit{functionality} they need in an example, and instead search for items that they associate with that functionality (e.g. I want a sprite to fly, so I search for ``airplane''). To help novices find an example based on these aesthetic properties, we might tag examples with relevant aesthetic tags, so that novices can find examples that include an airplane (or other flying object) when they search for it. We could also try to give feedback on search queries, e.g. ``Try searching for a verb -- what do you want the sprite(s) to do?''.

\vspace{1mm}

\noindent \textbf{Integration Barrier: }\textit{How do I integrate the example code into my own code?}

Our interview analysis identified \textbf{integration barriers} as the challenges students face when trying to integrate an example into their own code, after finding and opening it. For example, students noted differences between the example and their own code: \textit{``I may have looked at the `increase score' [example]. But I don't think I used that because I don't think we could have made it work... It wasn't a part of like our code..''} (P5). This difficulty integrating examples may be especially difficult in open-ended projects, where the students received examples that were distinct from the tasks they were trying to solve (i.e. it ``wasn't a part of'' their own code).

\textit{Low integration rate.} To understand how many examples students actually integrated into their projects, we investigated the 153 instances of opened example performed by the 17 students who opened examples. We treated an example that was revisited multiple times by a group as one distinct opened example, creating 77 distinct opened examples, covering all of the 48 examples we designed. We define \textbf{``integrated examples''} as the distinct opened example where students managed to use code from the example and integrate it into their program to create working code\footnote{This includes 2 examples that were successfully integrated and later deleted.}. We also define the \textbf{``integration rate''} as a measure of the proportion of opened examples that were ultimately integrated to  students' own code (i.e. \# integrated examples over \# opened examples). We found the integration rate over all opened examples to be (24.7\%) 19/77. This includes 9 times where students filled out the self-explanation prompt and copied the code to their own code, modifying it when needed. This number excludes 10 times where students attempted to integrate code but were unsuccessful.
 
We would not expect \textit{all} opened examples to be integrated into students code. For example, sometimes students \textit{browsed} examples, repeatedly opening examples in search of one they wanted. However, even when students searched for an example and found a relevant match, they did not often integrate it. For the 25 searched items that ended up retrieving at least one matched example, we found all of the top matched examples have been opened, but only 12\% (3/25) of them were later integrated to students projects. This suggests that students were encountering barriers to integration. Our thematic analysis revealed 4 specific \textit{types} of integration barriers that described how these difficulties occurred:  \textit{understanding}, \textit{mapping}, \textit{modification}, and \textit{testing} barriers, discussed below.

\vspace{1mm}
\noindent \textbf{Understanding Barrier: }\textit{How do I use an unfamiliar code block?}

\textbf{Understanding barriers} occurs when students encounter unfamiliar code blocks in an example, e.g., P2 found a ``glide'' block that they were unfamiliar with and asked \textit{``What is the glide [block]?''} (P2). In addition to not understanding a new API, students may also experience doubts about the usage of the API in the context of the example:  \textit{``I don't know how this will work with the broadcast start timer''} (P3), where ``broadcast'' is a code block that received ``start timer'' as its message. 
\begin{figure}
   \centering
   \includegraphics[width=.3\textwidth]{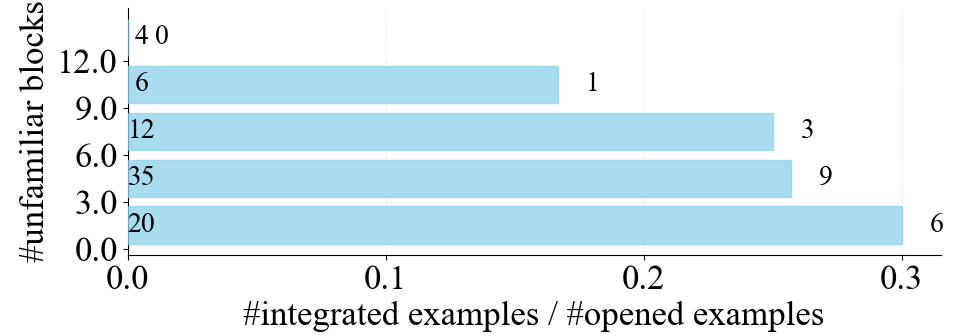} 
   \caption{\# unfamiliar blocks v.s. integration rate.}
   \label{fig:unfamiliar_type}
  \vspace{-3ex}
\end{figure}

The log data shows that the number of unfamiliar blocks indeed influenced students' ability to integrate an example into their own code. We used the number of distinct unfamiliar code blocks in each code example as the measure of unfamiliar blocks, and calculated it in the following way: 1) We took the set of blocks that appeared in at least 80\%\footnote{Other thresholds produced similar results.} student submissions in at least 1 of the 11 programming assignments prior to this open-ended project as a set of \textit{familiar blocks}. 2) In each example, A distinct code block that doesn't belong to the set of familiar block is an \textit{unfamiliar block}. Figure \ref{fig:unfamiliar_type} shows that the example integration rate continuously decreased from 30\% (6/20\footnote{6 integrated among 20 opened, shown by the right and left number in each bar}) to 0 (0/4), as the number of distinct unfamiliar blocks increased from 0-3 to 12-15. This shows that some students were unable to overcome the barrier of using unfamiliar blocks when the number of unfamiliar blocks increased in an example. 

\noindent\textit{Discussion. }
During open-ended programming, students can benefit from examples that demonstrate how to use features (e.g. blocks, APIs) that are unfamiliar, so it is important not to \textit{eliminate} unfamiliar code. Instead, our results show that students may find it difficult to understand examples when there are \textit{too many} new features (blocks) at once. Therefore, code example systems for novices may benefit from limiting the number of unfamiliar concepts to a certain threshold (e.g., 0-3 blocks). A system could also proactively show or link to documentation on unfamiliar concepts that students have likely not encountered before.

\vspace{1mm}
\noindent \textbf{Mapping Barrier: }\textit{How do I map a property of the example code to my own code?}

Students encountered\textbf{ mapping barriers} when trying to understand which parts of the example corresponded to existing parts of their own code, such as sprites --- P3 explained not knowing whether the example code should go into their current sprite or a new sprite: ``I don't know if creating another sprite is necessary.''; P5, when working with a multi-sprite example, had mistakenly copied example code to the wrong sprite, and later acknowledged that \textit{``Any difficulties that I might have had were... taking some time to understand how to change sprites to fit my project.''} (P5).

\begin{figure}
   \centering
   \includegraphics[width=.342\textwidth]{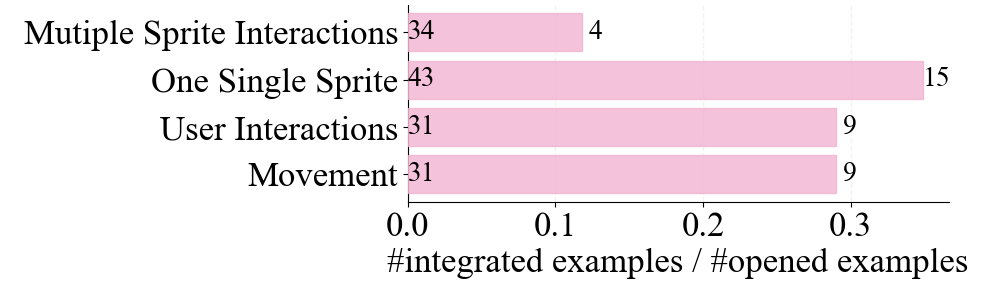} 
   \caption{example type v.s. integration rate}
   \label{fig:feature_type}
  \vspace{-3ex}
\end{figure}

Mapping barrier is also shown in students' challenges to integrate examples with multiple sprites. The 77 instances of opened examples included 4 primary categories: multiple-sprite interactions, single-sprite examples, user interactions, and movement. One example may belong to multiple categories. We calculated the integration rate for each category (shown in Figure \ref{fig:feature_type}), and found that multiple-sprite interaction examples, despite being the second-most popular category (with 34 distinct opens), had only an 11.8\% rate of successful integration, the \textit{lowest} among all other types of examples. This finding may be explained by the mapping barrier, since when integrating multiple-sprite examples, students face the two-fold barrier of finding ``which part of the example code completes my needed behavior'', as well as finding ``where in my code does the example go?''

One might argue that perhaps the challenge of integrating multiple-sprite examples may also be due to them being longer. We investigated students' ability to integrate examples as the size of the example grows. We divided examples into size bins (i.e. 1-25 blocks, 25-50 blocks, etc.) and compared multiple-sprites examples to other examples within each bin. We found the integration rate was always lower for multiple-sprite examples. For example, for examples of size 1-25 blocks, the integration rate was 13\% (3/23) for multiple-sprite examples v.s. 34\% (9/26) for others.
This shows that students struggle to integrate multiple-sprite examples to their code even when their sizes were small. However, students were still able to integrate some \textit{large} (50-75 block) single-sprite examples (60\%, 3/5). This suggests that mapping barriers with multiple-sprites, rather than an example's size, may explain students' challenges with integration.

\noindent\textit{Discussion. }
This difficulty in mapping an example's property to one's own code suggests that students need support to understand the example in the context of \textit{their own code}, e.g., potentially through adapting examples to match the student's current program. For Example Helper, this might mean changing the sprites in the example to match the student's, based on code similarity, or annotating when an example requires creating a new sprite.

\vspace{1mm}
\noindent \textbf{Modification Barrier: }\textit{How to modify the example code to fit my own needs?}

Modification barriers occurred when students were in the middle of or have completed integrating an example code into their own code, but encountered difficulties in modifying the example to what they actually needed. For example, P2 asked for an example to implement a bounce behavior. However, the example demonstrated how to bounce when hitting a sprite \textit{vertically}, while the students wanted to bounce after a \textit{horizontal} collision. The student gave up using the example because they were unable to modify the example to turn the correct number of degrees: \textit{``We were going to stick to what the code said, but the ball keeps falling off the paddle and we didn't know how to fix that, so I'm trying new stuff.''} (P2)

In our log data, we found 19 instances of example modification, which followed two distinct strategies: 1) \textit{build, test, modify, test} ($n=15$): students started by making code blocks based on example code, then tested and modified the examples by changing blocks. Among these students, 11 succeeded and kept the new code, while 4 were unsuccessful and removed the example code entirely. 2) \textit{modify while building} ($n=4$): students directly modified the example code as they constructed it (2/4 succeeded). Although with relatively high success rate (68 \%), some students who attempted to modify examples have been shown to have failed in doing so.

\noindent\textit{Discussion. }
Students' needs to modify the example show an active learning strategy \cite{chi2014icap}, which may cause the learner to mentally integrate the new information with their activated prior knowledge \cite{chi2014icap}. Because example code introduce a different context, and therefore not work correctly, students need \textit{debug} the examples through modification, which can be challenging \cite{sorvia2013notional}. We may therefore include options to toggle the example, or to encourage modification of a specific part of an example after they have used it in their program, which is also supported by the \textit{Use-Modify-Create} practice \cite{Lee2011}.

\noindent \textbf{Testing Barrier: } \textit{How to test the example code?}

Testing barriers occurred when students were expecting to test the example \textit{quickly}, but encounter difficulties in doing so. During our interview, two students asked the interviewer about how to test the example code \textit{immediately} after the student has opened the example (e.g.,  \textit{``Is there anywhere to see how the code actually works in the example?''} (P1)). In our log data, all 19 integrated examples were immediately tested once the students have completed making it. In addition, 9 opened example were tested in short time intervals, marked by at least two writing - testing cycles. This showed that students who managed to integrate the example code to their own code may have overcome the barrier of finding how to test the example code. However, our interview showed that some who were able to test the example code were still expecting quicker testing than what they experienced, and it's possible that students who did not integrate the examples successfully to their own code were discouraged by the difficulties of testing immediately. 

\textit{Discussion. }
Although our log data showed that all students who integrated examples to their own code have tried testing the example code by running it, students in the interviews were unsatisfied with the expectation that they have to first reconstruct the example in order to run it. Prior work has shown that when learning code examples, actually running the code and see how the code executes may lead to further reflections of the code itself \cite{wang2020iticse}. Our findings shows that just allowing students to view the animation next to the example code is insufficient; we should allow students easy access to run and test the example program directly. 

\noindent\textbf{Summary \& Discussion. }
We found evidence that students encounter decision, search and integration barriers, leading to lower levels of exploring, opening and using examples. Despite our focus on \textit{barriers}, our results still suggest that code examples have strong potential to support open-ended programming, as many students \textit{were} able to successfully integrate examples into their code.
Our results on learning barriers also show a strong connection between the challenges faced using examples, and more general programming skills, such as appropriate help-seeking \cite{marwan2020unproductive}, articulating what code does \cite{dorn2013lost}, and modifying code \cite{Lee2011}. In addition to design opportunities discussed above, our results also have implication for instructors, who often integrate examples into lectures and debugging sessions \cite{sorvia2013notional}, where students may still face each of the integration barriers we discussed.

\vspace{-1mm}
\section{Limitation \& Conclusions}
This work includes several limitations. With only five interviewees, our interview data may not generalize to other student groups. However, we validated our interview data with evidence from log data, showing some generalisability of these barriers. Additionally, some students programmed in pairs, others alone. We treated them equally as one unit of analysis, although they engage in different modes of programming. However, since the majority of our log analysis focused on each unit of example requests, and therefore the unit of analysis does not affect the validity of the data we reported. 

In conclusion, in this work, we presented the Example Helper system, which supports students' open-ended programming using code examples. We also identified students' learning barriers while using examples in open-ended programming, leading to design opportunities that may better support students.

\section{Acknowledgements}
This material is based upon work supported by the National Science Foundation under Grant No. 1917885.

\bibliographystyle{ACM-Reference-Format}
\balance
\bibliography{sample-base}


\end{document}